\documentclass[11pt,twoside]{article}
\usepackage{amssymb}

\usepackage{eurosym}

\usepackage{epsfig}
\usepackage{float}
\usepackage{amsmath,amssymb,amscd,amsthm,mathrsfs}
\usepackage{pstricks}
\usepackage{hyperref}

\hypersetup{colorlinks=true,linkcolor=blue,urlcolor=blue,citecolor=red}
\makeatletter

\@addtoreset{equation}{section}
\makeatother

\theoremstyle{plain}

\theoremstyle{definition}

\theoremstyle{remark}

\usepackage[top=2cm,bottom=2cm,left=2cm,right=2cm]{geometry}
\everymath{\displaystyle}
\begin{document}

\date{}
\title{A set of $q$-coherent states for the Rogers-Szeg\H{o} oscillator}
\author{ Zouha\"{i}r Mouayn$^{\ast,\sharp }$, Othmane El Moize$^{\flat }$   }
\maketitle
\vspace*{-0.7em}
\begin{center}
\textit{{\footnotesize ${}^{\ast }$ Department of Mathematics, Faculty of Sciences
and Technics (M'Ghila),\vspace*{-0.1em}\\ Sultan Moulay Slimane University, P.O. Box. 523, B\'{e}ni Mellal, Morocco \vspace*{0.2mm}\\[3pt]
${}^{\sharp}$ Department of Mathematics, KTH Royal Institute of Technology,\vspace*{-0.1em}\\
SE-10044, Stockholm, Sweden \vspace*{0.2mm}\\[3pt]
${}^{\flat }$ Department of Mathematics, Faculty of Sciences,\vspace*{-0.2em}\\Ibn Tofa\"{i}l University, P.O. Box. 133, K\'enitra, Morocco
 }
}
\end{center}
\thanks{}
\begin{abstract}
\scriptsize{We discuss a model of a $q$-harmonic oscillator based on Rogers-Szeg\H{o} functions. We combine these functions with a class of   $q$-analogs of complex Hermite polynomials  to construct a new set of coherent states depending on a nonnegative integer  parameter $m$. Our construction leads to a new $q$-deformation of the $m$-true-polyanalytic Bargmann transform whose range defines a generalization of the Arik-Coon space. We also give an explicit formula for  the reproducing kernel of this space.  The obtained results  may be exploited to define a $q$-deformation of the Ginibre-$m$-type process on the complex plane. 
}
\end{abstract}
\date{\today}
\section{Introduction}
Coherent states (CS) are a type of quantum states which were first introduced
by Schr\"{o}dinger \cite{schro} when he described certain states of the harmonic oscillator (HO). Later, 
Glauber \cite{Glauber} used these states  for his quantum mechanical description of coherent
laser light, and coined the term CS. Like canonical CS, they satisfy
a set of relevant properties and they too have found wide applications in different branches of physics such as quantum optics, statistical
mechanics, nuclear physics and condensed matter physics \cite{KLSK85}. One of many definitions of CS is a special superposition with the form
\begin{equation}
\Psi _{z}:=\left( e^{z\bar{z}}\right) ^{-1/2}\sum_{j=0}^{+\infty }\frac{\bar{z%
}^{j}}{\sqrt{j!}}\phi _{j},  \label{CS01}
\end{equation}%
where the $\phi_j$'s span a Hilbert space $\mathscr{H}$ usually called Fock space.

The terminology of generalized CS (GCS) was first appeared and studied in \cite{Birula,Stoler} in connection with states discussed in \cite{TG65}. These states are usually associated with potential algebras other than the
oscillator one \cite{KLSK85, pere86, Zhan90}. An important example is provided by the $q$%
-deformed CS ($q$-CS for brevity) related to deformations of boson
operators \cite{ACO, Bi89, Mac89}. $q$-CS are usually constructed in a way that
they reduce to their standard counterparts as $q\rightarrow 1$ and are 
special case of  a more general class, the so-called $f$-deformed CS \cite{MMSZ}.  Among $q$-CS, there are those associated with the relation 
\begin{equation}\label{qcommu}
a_{q}a_{q}^{\ast }-qa_{q}^{\ast }a_{q}=1
\end{equation}%
with $0<q<1$, where  $a_{q}$ are often termed maths-type $q$-bosons operators \cite%
{Solo94, JKAS94} because the\textit{\ basic }numbers and special functions attached
to them have been extensively studied in  mathematics \cite{GR}. As a consequence, the corresponding wavefunctions of  these $q$-boson operators were found in terms of many orthogonal $q$-polynomials \cite{atakall,AS93,AS932}.

 The $q$-CS may be defined \cite{ACO} through a $q$-analog of the expansion \eqref{CS01} by a superposition of a set of number states $\phi _{j}^{(q)}$  spaning  a Hilbert space $\mathscr{H}_{q}$ which stands for a $q$-analog of the Fock space $\mathscr{H}$ as :
\begin{equation}  \label{CS02}
\Psi _{z}^{q}:=\left( e_{q}(z\bar{z})\right) ^{-\frac{1}{2}%
}\sum_{j=0}^{+\infty }\frac{\bar{z}^{j}}{\sqrt{[j]_{q}!}}\phi _{j}^{(q)},
\end{equation}%
for $z\in\mathbb{C}_q:=\{\zeta\in\mathbb{C},\:(1-q)\zeta\bar{\zeta}<1\}$ where $\lbrack j]_{q}=\frac{1-q^{j}}{1-q} \rightarrow j$ as $q\rightarrow 1$,
\begin{equation}\label{qfactor}
[j]_q!=\frac{(q;q)_j}{(1-q)^j}
\end{equation}
denotes the $q$-factorial of  $j$ and 
\begin{equation}\label{qexpo1}
e_q(\xi):= \displaystyle\sum_{n\geq 0}  \frac{\xi^n}{[n]_q!}=\frac{1}{((1-q)\xi;q)_{\infty}},\quad|\xi|<\frac{1}{1-q}
\end{equation}
 being a $q$-exponential function satisfying  $e_{q}(\xi)\to e^{\xi}$ as $q\to 1$.  For  details on basic notations of $q$-calculus we refer to \cite{GR, KS, Ism29}.

In this paper, we adopt the \textit{Hilbertian probabilistic} formalism \cite%
{gazeau} in order to generalize the expression in \eqref{CS02} with respect to a fixed integer
parameter $m\in \mathbb{N}$. Precisely, we here introduce the following superposition of the $\phi_j^{(q)}$'s: 
\begin{equation}
\Psi _{z}^{q,m}:=\left( \mathcal{N}_{q,m}(z\bar{z})\right) ^{-\frac{1}{2}%
}\sum_{j=0}^{+\infty }\overline{\Phi_{j}^{q,m}(z)}\phi _{j}^{(q)},  \label{CS03}
\end{equation}%
where $\mathcal{N}_{q,m}(z\bar{z})$ is a normalization factor (given by \eqref{normalization} below) and 
\begin{equation}\label{Cjmq}
\Phi_{j}^{q,m}(z):=\frac{q^{\binom{m\wedge j}{2}}\sqrt{1-q}%
^{|m-j|}|z|^{|m-j|}e^{-i(m-j)arg(z)}}{((q;q)_{m\vee j})^{-1}(-1)^{m\wedge j}(q;q)_{|m-j|}\sqrt{%
q^{mj}(q;q)_{m}(q;q)_{j}}}P_{m\wedge j}\left( (1-q)z\bar{z}%
;q^{|m-j|}|q\right)
\end{equation}%
are coefficients  defined in terms of Wall polynomials (\cite{KS}, p.109) :
\begin{eqnarray}\label{Wall}
   P_n(x;a|q)=\setlength\arraycolsep{1pt}
{}_2 \phi_1\left(\begin{matrix}q^{-n},0 \\ aq \end{matrix}\left|q;qx\right.\right).
  \end{eqnarray}
   Here, $m\vee j=\max (m,j)$, $m\wedge j=\min (m,j)$ and ${}_2 \phi_1$ is the basic hypergeometric series (\cite{KS}, p.12). In the case  $m=0$, $\Phi_{j}^{q,0}(z)$
reduce to the coefficient $z^{j}/\sqrt{[j]_{q}!}$. For general $m\in \mathbb{N},$ the coefficients \eqref{Cjmq} are a slight modification of
a class of $2D$ orthogonal $q$-polynomials, denoted $H_{j,m}\left( z,\overline{z}%
|q\right) ,$ which are  $q$-deformation \cite{IZ} of the well
known $2D$ complex Hermite polynomials \cite{IK} $H_{j,m}\left( z,\overline{z}\right)$  whose Rodriguez-type formula reads  $%
H_{j,m}\left( z,\overline{z}\right)=\left( -1\right) ^{j+m}e^{z%
\overline{z}}\partial _{z}^{j}\partial _{\overline{z}}^{m}e^{-z\overline{z}}$%
. This last expression turns out to
be the \textit{diagonal representation} (or  upper symbol \cite{Mehta} with respect to CS \eqref{CS01}) of the operator $\left( a^{\ast }\right) ^{j}a^{m}$  whose expectation values are needed \cite{quesne02} in the study of squeezing properties  involving the Heisenberg uncertainty relation. Here, $a$ and $a^{\ast}$ are the classical annihilation and creation operators. In other words, the polynomial $H_{j,m}\left( z,\overline{z}\right) $  also represents a classical
observable on the phase space or a Glauber-Sudarshan $P$-function for $\left( a^{\ast }\right) ^{j}a^{m}$. This  means that the G$q$-CS $\Psi _{z}^{q,m}$ in \eqref{CS03} we are introducing may play an analog  role in the Berezin \textit{de}-quantization procedure of $\left( a_{q}^{\ast }\right) ^{j}a_{q}^{m},$ see \cite{B91} for a similar discussion related the $q$-Heisenberg-Weyl group.

 Here, we introduce an explicit realization of $q$-creation and $q$-annihilation operators associated to a Rogers-Szeg\H {o} oscillator whose eigenstates   are chosen to be our vectors $\phi _{j}^{(q)}$. Next, we give a closed form for the superposition \eqref{CS03} and the associated CS transform (CST) $\mathcal{B}_m^{(q)}$ which may be viewed as a new $q$-deformation of the $m$-true polyanalytic Bargmann transform \cite{Mouayn1,Cras}. As a consequence,  the range of $\mathcal{B}_m^{(q)}$  defines a generalization of the well known  Arik-Coon space \cite{ACO}, whose  reproducing kernel will be given explicitly. The latter one may be used to  define a $q$-deformation of the Ginibre-$m$-type process on the complex plane \cite{shirai}.

The paper is organized as follows.  In Section 2, we give a brief review of  the standard coherent state formalism. In section 3, we discuss the Hilbert space carrying the coefficients needed in the superposition defining our G$q$-CS.  Section 4 is devoted to  discuss a model of a $q$-deformed HO based on Rogers-Szeg\H {o} functions with a realization of $q$-creation and $q$-annihilation operators on $L^2(\mathbb{R})$. In Section 5, we introduce a new set of G$q$-CS for the Rogers-Szeg\H {o} oscillator and we give an explicit formula for the associated CST whose range provides us with a generalization of the Arik-Coon space. For the latter one we also obtain explicitly the reproducing kernel. Most technical proofs and calculations are postponed in Appendices.
\section{A CS formalism }\label{CSsection}
Here, we adopt the prototypical model of CS presented in (\cite{gazeau}, pp.72-77) and described as follows. Let $X$ be a set equipped with a measure $d\mu$ and $L^2(X)$ the Hilbert space of $d\mu$-square integrable functions
$f(x)$ on $X$. Let $\mathcal{A}^2\subset L^2(X)$ be a closed subspace with an orthonormal basis $\lbrace\Phi_k\rbrace_{k=0}^{\infty}$ such that 
\begin{equation}
\mathcal{N}(x):=\displaystyle\sum_{j\geq 0} |\Phi_j(x)|^2<+\infty,\qquad x\in X.
\end{equation}
Let $\mathscr{H}$ be another (functional) Hilbert space with  the same dimension as $ \mathcal{A}^2$ and $\lbrace\phi_k\rbrace_{k=0}^{\infty}$
is a given orthonormal basis of  $\mathscr{H}$.  Then, consider the family of states $\lbrace\Psi_x\rbrace_{x\in X}$ in $\mathscr{H}$, through the following linear superpositions
\begin{equation}\label{CSdef11}
\Psi_x:=(\mathcal{N}(x))^{-1/2}\displaystyle\sum_{j\geq 0} \Phi_j(x)\phi_j.
\end{equation}
These CS obey the normalization condition
\begin{equation}
\langle \Psi_x|\Psi_x\rangle_{\mathscr{H}}=1
\end{equation}
and the following resolution of the identity operator on $\mathscr{H}$
\begin{equation}\label{Resofidd}
\textbf{1}_{\mathscr{H}}=\int_X |\Psi_x\rangle\langle \Psi_x |\mathcal{N}(x) d\mu(x)
\end{equation}
which is expressed in terms of Dirac’s bra-ket notation $|\Psi_x\rangle\langle\Psi_x |$ meaning  the rank one operator defined by $\varphi\mapsto \langle \Psi_x|\varphi \rangle_{\mathscr{H}}\cdot\Psi_x$, $\varphi \in \mathscr{H}$. The choice of the Hilbert space $\mathscr{H}$ defines in fact a quantization of the space $X$ by
the coherent states in \eqref{CSdef11}, via the inclusion map $X\ni x\mapsto \Psi_x \in \mathscr{H}$ and the property
\eqref{Resofidd} is crucial in setting the bridge between the classical and the quantum worlds. The CST associated with the set $\Psi_x$  is the  map  $\mathcal{B}:\mathscr{H}\longrightarrow \mathcal{A}^2$ defined for every $x\in X$ by 
\begin{equation}\label{defbargtrans}
\mathcal{B}[\phi](x):=\left(\mathcal{N}(x)\right)^{1/2}\langle\phi,\Psi_x\rangle_{\mathscr{H}}.
\end{equation}
From the resolution of the identity \eqref{Resofidd} and  for $\phi,\,\psi\in \mathscr{H}$, we have $\langle\phi,\psi\rangle_{\mathscr{H}}=\langle\mathcal{B}[\phi],\mathcal{B}[\psi]\rangle_{L^2(X)}$ 
meaning that $\mathcal{B}$ is an isometric map.

The formula \eqref{CSdef11} can be considered as a generalization of the series expansion of the
canonical CS in \eqref{CS01} with $\lbrace\phi_j\rbrace_{j=0}^{\infty}$ being an orthonormal basis of the Hilbert space $\mathscr{H}:=L^2(\mathbb{R})$, consisting of  eigenstates of the quantum HO given by
\begin{equation}
\phi_j(\xi)=\left(\sqrt{\pi} 2^j j!\right)^{-1/2}H_j(\xi)e^{-\frac{1}{2}\xi^2},\quad \xi\in\mathbb{R}
\end{equation}
where
\begin{equation}\label{Hermitdef}
H_j(\xi):=j!\displaystyle\sum_{k=0}^{\lfloor j/2\rfloor}\frac{(-1)^k}{k!(j-2k)!}(2\xi)^{j-2k}
\end{equation}
is the Hermite polynomial of degree $j$  (\cite{KS}, p.59). Here, the space $\mathcal{A}^2$ is the Bargmann-Fock space $\mathfrak{F}(\mathbb{C})$ of entire complex-valued functions which are $\pi^{-1}e^{-z\bar{z}}d\lambda$-square integrable, $\mathcal{N}(z)=e^{z\bar{z}}$, $z\in\mathbb{C}$ and  $d\lambda$ denotes the Lebesgue measure on $\mathbb{C}\cong \mathbb{R}^2$. In this case, the associated CST   
$\mathcal{B}_0:L^2(\mathbb{R})\rightarrow \mathfrak{F}(\mathbb{C})$, defined for any function $f\in L^2(\mathbb{R})$ by 
\begin{equation}\label{barcl}
\mathcal{B}_0[f](z):=\pi^{-\tfrac{1}{4}}\int_{\mathbb{R}}e^{ -\tfrac{1}{2}\xi^2+\sqrt{2}\xi z-\tfrac{1}{2}z^2}f(\xi)d\xi,\quad z\in\mathbb{C}
\end{equation}
turns out to be the well known Bargmann transform (\cite{B}, p.12).  
\section{The Hilbert space $\mathfrak{F}_q$}
 We observe that the coefficients \vspace*{-1em}
\begin{equation}
\Phi_{j}^{q}(z)=\frac{z^{j}}{\sqrt{[j]_{q}!}},\quad j=0,1,2,\ldots ,
\label{qcoeffi}
\end{equation}%
occurring in the number state expansion of $\Psi _{z}^{q}$ in \eqref{CS02}  constitute a particular case of the complex-valued functions $\Phi_{j}^{q,m}(z)$ in \eqref{Cjmq}.  Indeed, for $m=0$ we have that $%
\Phi_{j}^{q,0}(z)=\Phi_{j}^{q}(z)$. For $m\in \mathbb{N}$, these coefficients are a slight modification of the $2D$ complex
orthogonal polynomials (\cite{IZ}, p.6783) :

\begin{equation}\label{Hmjqdef}
H_{m,j}(z,\zeta |q):=\displaystyle\sum_{k=0}^{m\wedge j}%
\begin{bmatrix}
m \\ 
k%
\end{bmatrix}%
_{q}%
\begin{bmatrix}
j \\ 
k%
\end{bmatrix}%
_{q}(-1)^{k}q^{\binom{k}{2}}(q;q)_{k}z^{m-k}\zeta ^{j-k},\ z,\zeta \in 
\mathbb{C}.
\end{equation}%
Precisely,
\begin{equation}\label{phi-vs-Hmj}
\Phi_{j}^{q,m}(z):=\frac{H_{m,j}(\sqrt{1-q}z,\sqrt{1-q}\bar{z}|q)}{\sqrt{%
q^{mj}(q;q)_{j}(q;q)_{m}}}.
\end{equation}%
Note, also, that by using \eqref{qfactor}, one can check that $\Phi _{j}^{q,m}(z)\rightarrow\left( m!j!\right)^{-1/2}H_{m,j}(z,\bar{z})$ as $q\to 1$, where 
\begin{equation}
H_{m,j}(z,\zeta )=\sum_{k=0}^{m\wedge j}(-1)^{k}k!\binom{m}{k}\binom{j}{k}%
z^{m-k}\zeta ^{j-k}
\end{equation}%
denote the $2D$ complex Hermite polynomials introduced by It\^{o} \cite%
{IK} in the context of complex Markov processes.

Furthermore, we observe that the functions $\Phi_{j}^{q,m}(z)$ constitute an orthonormal system in the Hilbert space $\mathfrak{F}_{q}:=L^2(\mathbb{C}_{q,m}, d\mu_q(z))$  of square integrable functions on $\mathbb{C}_{q,m}:=\{\zeta\in\mathbb{C},\:(1-q)\zeta\bar{\zeta}<q^m\}$ with respect  to the measure
\begin{equation}
d\mu _{q}(z)=\frac{d\theta }{2\pi }\otimes \sum_{l\geq 0}\frac{%
q^{l}(q;q)_{\infty }}{(q;q)_{l}}\delta(r-\frac{q^{%
\tfrac{1}{2}l}}{\sqrt{1-q}}),  \label{dmuq}
\end{equation}%
$z=re^{i\theta},\ r\in\mathbb{R}^+$, $\theta\in[0,2\pi)$ and $\delta(r-\bullet)$ is the Dirac mass function at the point $\bullet$. Indeed, we may assume that $m\geq j$ because of the symmetry $H_{r,s}(z,w|q)=H_{s,r}(w,z|q)$, then we use the expression \eqref{Cjmq} to obtain
\begin{eqnarray*}
\langle \Phi_j^{q,m}, \Phi_k^{q,m} \rangle_{\mathfrak{F}_{q}}&=& \displaystyle\int _{\mathbb{C}_{q,m}}\Phi_j^{q,m}(z)\overline{\Phi_k^{q,m}}(z)d\mu_{q}(z)\cr
&=&(-1)^{j+k}\frac{(q;q)_mq^{\binom{j}{2}}\sqrt{1-q}^{m-j}}{(q;q)_{m-j}(q;q)_m\sqrt{q^{mj}(q;q)_j}}\frac{(q;q)_m q^{\binom{k}{2}}\sqrt{1-q}^{m-k}}{(q;q)_{m-k}\sqrt{q^{mk}(q;q)_k}}\displaystyle\int _{0}^{2\pi} e^{i\theta(j-k)}\frac{d\theta}{2\pi}\cr
&\times&\displaystyle\sum_{l=0}^{\infty}\frac{q^l(q;q)_{\infty }}{(q;q)_l}P_j(r^2,q^{m-j}|q)P_k(r^2,q^{m-k}|q) r^{2m-j-k}\delta(r-\frac{q^{\tfrac{l}{2}}}{\sqrt{1-q}})
\end{eqnarray*}
\begin{equation}\label{coefortho}
=\frac{(-1)^{j+k}(q;q)_mq^{\binom{j}{2}+\binom{k}{2}-mj}}{(q;q)_j(q;q)_{m-j}(q;q)_{m-k}}\delta_{j,k}\; \displaystyle\sum_{l=0}^{\infty}\frac{q^{l(1+m-j)}(q;q)_{\infty }}{(q;q)_l}P_j(q^l,q^{m-j}|q)P_k(q^l,q^{m-j}|q)=\delta_{j,k}.
\end{equation}
Here,  we have used  the orthogonality relations of  Wall polynomials (\cite{KS}, p.107):
\begin{equation}\label{wall-orthogo}
\sum_{l=0}^{\infty} \frac{(\tau q)^l}{(q;q)_l}P_s(q^l;\tau|q)P_n(q^l;\tau|q)=\frac{(\tau q)^n}{(\tau q;q)_{\infty}}\frac{(q;q)_n}{(\tau q;q)_n}\ \delta_{s,n},\quad 0<\tau<q^{-1}
\end{equation}
for  parameters $\tau=q^{m-j}$, $s=j$ and $n=k$.

Finally,  we denote by $\mathcal{A}^{2}_q(\mathbb{C}%
) $ the completed space of holomorphic functions on $\mathbb{C}_{q}=\mathbb{C}_{q,0}$,
equipped with the scalar product 
\begin{equation}
\langle \varphi ,\phi \rangle =\int_{\mathbb{C}_{q}}{\varphi (z)}\overline{%
\phi (z)}d\mu _{q}(z).
\end{equation}%
 The element $\Phi_{j}^{q,0}(z)=([j]_{q}!)^{-1/2}z^{j}$ form an orthonormal
basis of the space $\mathcal{A}^{2}_q(\mathbb{C})$ which co\"{\i}ncides
with Arik-Coon space \cite{ACO} whose reproducing kernel is given by 
\begin{equation}
K_q(z,w):=e_q(z\bar{w}), \qquad z,w\in\mathbb{C}_q.
\end{equation}
Note that by letting $q\rightarrow 1$, the measure $d\mu _{q}$ reduces to the Gaussian measure $\pi ^{-1}e^{-z\bar{z}}d\lambda $ on $\mathbb{C}$. 
\section{A Rogers-Szeg\H {o} Hamiltonian} 
Following \cite{Atak}, we consider a $q$-deformed creation and annihilation operators as follows:
\begin{eqnarray}
B^{\ast}_q&=&\frac{e^{i\kappa x}}{i\sqrt{1-q}}\left(e^{i\kappa x}-q^{3/4}e^{i\kappa\partial_x}\right)\label{creation}\\
B_q&=&\frac{-e^{-i\kappa x}}{i\sqrt{1-q}}\left(e^{-i\kappa x}-q^{1/4}e^{i\kappa\partial_x}\right).\label{annih}
\end{eqnarray}
Here $\kappa$ is a deformation parameter related to a finite-difference method with
respect to $x$ and  $q=e^{-2\kappa^2},\:\kappa>0$. Note that for $a\in\mathbb{C}$ the operator $e^{a\partial_x}$ acts on a function $f(x)$ as   $e^{a\partial_x}[f](x)=f(x+a).$ Moreover, it is not difficult to verify that  operators \eqref{creation}-\eqref{annih} satisfy the $q$-commutation relation 
\begin{equation}\label{qcommu}
[B_q,B^{\ast}_q]=B_qB^{\ast}_q-q B^{\ast}_qB_q=1.
\end{equation}
Similarly to the case of the quantum  HO, the $q$-deformed HO is described by the Hamiltonian
\begin{equation}\label{hamiltionian}
H_q=\frac{1}{2}\hbar\omega\left(B_qB^{\ast}_q+B^{\ast}_qB_q\right)
\end{equation}
where $\omega$ is the oscillator frequency and $\hbar$ denotes the Planck’s constant. Explicitly,
\begin{equation}\label{hami-expli}
H_q=\frac{1}{2(q-1)}\hbar \omega\left(-2+(q^{1/4}+q^{5/4})e^{i\kappa x}e^{i\kappa\partial_x } +(q^{-1/4}+q^{3/4})e^{-i\kappa x}e^{i\kappa\partial_x }+(q^{3/2}+q^{1/2})e^{2i\kappa\partial_x }\right).
\end{equation} 
For the sake of simplicity we will take $\hbar= \omega=1$. The eigenstates of  the Hamiltonian in \eqref{hami-expli} are given by 
\begin{equation}\label{RSfunc}
\varphi _j^{RS}(x):=\frac{(i\sqrt{q})^j}{\pi^{\frac{1}{4}}\sqrt{(q;q)_j}}H_j\left(-e^{2i\kappa x};q\right)e^{-\frac{1}{2}x^2},\; x\in\mathbb{R}
\end{equation}
in terms of the  Rogers–Szeg\H{o} polynomials (\cite{Ism29}, p.455)
\begin{equation}\label{RG-poly}
H_{n}(\xi;q):=\sum_{k=0}^{n}\left[ 
\begin{array}{c}
n \\ 
k%
\end{array}%
\right] _{q}(q^{-1/2}\xi)^{k},\quad \xi\in\mathbb{C}.
\end{equation}
It was proved  (\cite{Atak}, p.612)  that the functions $\varphi _j^{RS}$ satisfy the orthonormality relations on the full real line, i.e.,
\begin{equation}\label{arthogphij}
\int\limits_{\mathbb{R}}\varphi _j^{RS}(x)\varphi _k^{RS}(x)dx=\delta _{jk}.
\end{equation}

Furthermore, each of these functions can also be reproduced by
iterating $j$ times the action of the $q$-creation operator $B_q^{*}$ on the Gaussian function as 
\begin{equation}
\varphi _j^{RS}(x)=\left( B_{q}^{\ast}\right) ^{j}\,\left( \pi ^{%
\frac{-1}{4}}e^{\tfrac{-1}{2}x ^{2}}\right)
\end{equation}
and one can check  that the  operators $B_q^{\ast}$ and $B_{q}$ act on them as
\begin{equation}\label{actionBB}
B^{\ast}_q\varphi _j^{RS}(x)=\varphi _{j+1}^{RS}(x)\quad, \qquad B_q\varphi _j^{RS}(x)=[j]_q\varphi _{j-1}^{RS}(x). 
\end{equation}
Then, it follows that  the  Hamiltonian  \eqref{hamiltionian} is diagonal on these states  and has the eigenvalues 
\begin{equation}
\varepsilon_j^q=\frac{1}{2}\left([j+1]_q+[j]_q\right)
\end{equation}
as $q$-deformed energy levels. 

Finally, let us mention that in the limit $q\to 1$ ($\kappa\to 0$), the operators \eqref{creation}-\eqref{annih} reduce to the well-known non relativistic HO creation and annihilation operators. Also,  from  (\cite{Atak}, p.614) one  has that 
\begin{equation}\label{HOef}
\displaystyle\lim_{q\rightarrow 1} \varphi _j^{RS}(x)=\left(\sqrt{\pi} 2^j j!\right)^{-1/2}H_j(x)e^{-\frac{1}{2}x^2}
\end{equation}
where $H_j(\cdot)$ is the Hermite polynomial \eqref{Hermitdef}, meaning that the $\varphi _j^{RS}$'s  stand for   $q$-analogs of the HO wave functions which justify our choice in \eqref{RSfunc}.
\section{A new set of G$q$-CS} 
For $m\in\mathbb{N}$ and $q\in]0,1[$, we define a new set of G$q$-CS by the  following superposition of the $\varphi_j^{RS}$'s:
\begin{equation}\label{Eq:phizqm}
\Psi_{z}^{q,m} :=(\mathcal{N}_{q,m}(z\bar{z}))^{-\tfrac{1}{2}}\sum_{j\geq 0}\overline{\Phi_j^{q,m}(z)}\varphi_j^{RS}
\end{equation}
where
\begin{equation}\label{normalization}
\mathcal{N}_{q,m}(z\bar{z})=\frac{q^{-m}(q^{1-m}(1-q)z\bar{z};q)_m }{(q^{-m}(1-q)z\bar{z};q)_{\infty}}  
\end{equation}
is the  normalization factor ensuring $\langle \Psi_{z}^{q,m}|\Psi_{z}^{q,m}\rangle_{L^2(\mathbb{R})}=1$, which is well defined for $z\in\mathbb{C}_{q,m}$.

Using the normalization factor \eqref{normalization} together with the orthonormality relations \eqref{coefortho} of the coefficients $\Phi_j^{q,m}(z)$ and the completeness of the basis $\varphi_j^{RS}$ in $L^2(\mathbb{R})$ one can check that the states  $\Psi_{z}^{q,m}$   provide us with a resolution of the identity operator  on $L^2(\mathbb{R})$ as 
\begin{equation}\label{operiden}
\textbf{1}_{L^2(\mathbb{R})}=\int_{\mathbb{C}}\mathcal{N}_{q,m}(z\bar{z})d\mu_{q}(z)|\Psi _{z}^{q,m}\rangle \langle\Psi _{z}^{q,m} |.
\end{equation}

We now establish (see Appendix A for the proof) that the wave function of the CS \eqref{Eq:phizqm} can be expressed as 
\begin{eqnarray}\label{SalCHi}
\Psi_{z}^{q,m}(\xi) &=&(-1)^m\left(\frac{q^m\,e^{-\xi^2}(q^{-m}(1-q)z\bar{z};q)_{\infty}}{\sqrt{\pi}(q;q)_m(q^{1-m}(1-q)z\bar{z};q)_m}\right)^{\tfrac{1}{2}} \frac{q^{-m/4}e^{im\kappa \xi}}{(iz\sqrt{\tfrac{1-q}{q^{m-1}}},-iz \sqrt{\tfrac{1-q}{q^m}} e^{2i\kappa \xi} ;q)_{\infty}}\cr
&\times& 
Q_m\left(\frac{ie^{-i\kappa\xi}q^{1/4}-ie^{i\kappa\xi}q^{-1/4}}{2};zq^{-1/4}\sqrt{\tfrac{1-q}{q^{m-1}}e^{i\kappa\xi}},\bar{z}q^{1/4}\sqrt{\tfrac{1-q}{q^{m-1}}e^{-i\kappa\xi}}\:\;|q  \right),  \quad\xi\in\mathbb{R},
\end{eqnarray}
in terms of   AL-Salam-Chihara polynomials $Q_m(.;.,.|q)$, which are defined by (\cite{KS}, p.80):
\begin{equation}\label{SaChidef}
Q_m(x;\alpha,\beta|q)=\frac{(\alpha \beta;q)_m}{\alpha^m}\:_3 \phi_2\left(\begin{matrix}q^{-m},\alpha u,\alpha u^{-1} \\ \alpha \beta,0 \end{matrix}\Big|q;q\right),\qquad x=\frac{1}{2}(u+u^{-1}).
\end{equation}
Further, by applying  \eqref{defbargtrans}, the CST $\mathcal{B}_{m}^{(q)}:L^{2}(\mathbb{R}%
)\longrightarrow L^2(\mathbb{C}_{q,m}, d\mu_q)$ defined  by 
\begin{equation}\label{CSTdef}
\mathcal{B}_{m}^{(q)}[f](z)=(\mathcal{N}_{q,m}(z\bar{z}))^{\tfrac{1}{2}}\langle f,\,\Psi _{z}^{q,m}\rangle_{L^2(\mathbb{R})},\;z\in \mathbb{C}_{q,m},
\end{equation}
is an isometric map. Explicitly, 
\begin{equation*}
\mathcal{B}_{m}^{(q)}[f](z)=\left(\frac{q^{-m/2}}{\sqrt{\pi}(q;q)_m}\right)^{\tfrac{1}{2}}\frac{(-1)^m}{(iz\sqrt{\tfrac{1-q}{q^{m-1}}};q)_{\infty}} 
\end{equation*}
\begin{equation}\label{CSBMQ}
\times \int_{\mathbb{R}}\frac{e^{im\kappa \xi} e^{-\frac{1}{2}\xi^2} }{(-iz \sqrt{\tfrac{1-q}{q^m}} e^{2i\kappa \xi};q)_\infty}Q_m\left(\frac{ie^{-i\kappa\xi}q^{1/4}-ie^{i\kappa\xi}q^{-1/4}}{2};zq^{-1/4}\sqrt{\tfrac{1-q}{q^{m-1}}e^{i\kappa\xi}},\bar{z}q^{1/4}\sqrt{\tfrac{1-q}{q^{m-1}}e^{-i\kappa\xi}}|q  \right)
 f(\xi)d\xi
\end{equation}
for every  $z\in \mathbb{C}_{q,m}$.\medskip\\

Particularly, for $m=0$, the wave function  \eqref{SalCHi} has the form
\begin{equation}\label{CS2}
\Psi_{z}^{q,0}(\xi)   = (e_q(z\bar{z}))^{-\tfrac{1}{2}} \frac{\pi^{-\frac{1}{4}}}{(-iz\sqrt{ 1-q } e^{2i\kappa \xi} ;q)_{\infty}} \frac{e^{-\frac{1}{2}\xi^2}}{(iz\sqrt{q(1-q)}  ;q)_{\infty}}, \quad \xi\in \mathbb{R}
\end{equation}
where   $z\in \mathbb{C}_{q} :=\mathbb{C}_{q,0}$ which is the domain of  convergence of the exponential function in $e_q(\cdot)$. In this case, the transform \eqref{CSBMQ} reduces to $\mathcal{B}_{0}^{(q)}:L^{2}(\mathbb{R}%
)\longrightarrow \mathcal{A}_q^2(\mathbb{C})$
\begin{equation}\label{B0q}
\mathcal{B}_0^{(q)}[f](z)=\frac{ \pi^{-\frac{1}{4}}}{ 
(iz\sqrt{q(1-q)} ;q)_{\infty}} \int_{\mathbb{R}}\frac{e^{-\frac{1}{2}\xi^2} }{(-iz \sqrt{ 1-q}  e^{2i\kappa \xi};q)_\infty}
 f(\xi)d\xi
\end{equation}
for every $z\in\mathbb{C}_q$. Moreover, when $q \to 1 $, $\mathcal{B}_{0}^{(q)}$ goes to the  Bargmann transform \eqref{barcl}.

In addition, by letting $q \to 1$, $\mathcal{B}_{m}^{(q)}$ goes to the generalized Bargmann transform  
$\mathcal{B}_m:L^2(\mathbb{R})\rightarrow \mathcal{A}^2_m(\mathbb{C})\subset L^2(\mathbb{C},\pi^{-1}e^{-z\bar{z}}d\lambda)$, defined by \cite{Mouayn1}:
\begin{equation}\label{tr1}
\mathcal{B}_m[f](z)=(-1)^m(2^mm!\sqrt{\pi})^{-\tfrac{1}{2}}\int_{\mathbb{R}}e^{-\tfrac{1}{2}z^2-\tfrac{1}{2}\xi^2+\sqrt{2}\xi z}H_m\left( \xi-\frac{z+\bar{z}}{2}\right)f(\xi)d\xi
\end{equation}
where $H_m(.)$ denotes the Hermite polynomial, see Appendix B for the proof. This transform, also called  $m$-true-polyanalytic Bargmann transform, has found applications in time-frequency analysis \cite%
{Abr2010} and  discrete quantum dynamics \cite{AoP}. For more details on \eqref{tr1}, see \cite{AF} and reference therein. Here, the arrival space $\mathcal{A}^2_m(\mathbb{C})$ is the generalized Bargmann-Fock space whose reproducing kernel is given by \cite{AIM}:
\begin{equation}
K_m(z,w)=e^{z\bar{w}}L_m^{(0)}\left(|z-w|^2\right),\quad z,w\in\mathbb{C}.
\end{equation}

Consequently, the range of the CST \eqref{CSBMQ}  leads us to  define a generalization, with respect to $m\in\mathbb{N}$,  of the Arik-Coon space $\mathcal{A}_q^2$ as $\mathcal{A}_{q,m}^2=\mathcal{B}_{m}^{(q)}(L^2(\mathbb{R}))$ which also coincides with the closure in $\mathfrak{F}_q$ of the linear span of $\lbrace \Phi_j^{q,m}\rbrace_{j\geq 0}$. In Appendix C, we prove that the reproducing kernel of the space $\mathcal{A}_{q,m}^2$ is given by
\begin{eqnarray}\label{kern11}
 K_{q,m}(z,w)=\frac{(q^{1-m}(1-q)z\bar{z};q)_{m}}{ q^m(q^{-m} (1-q)z\bar{w};q)_\infty}
{}_3\phi_2\left(\begin{array}{c}q^{-m},\frac{z}{w},q\frac{\bar{z}}{\bar{w}}\\
 q,q^{-m+1}(1-q)z\bar{z}\end{array}\Big|q;q(1-q)w\bar{w}\right)
\end{eqnarray}
for every $z,w\in \mathbb{C}_{q,m}$. In particular, the limit $K_{q,m}(z,w)\to K_{m}(z,w)$ as $q\to 1$ holds true.\medskip\\
Finally, note that the expression \eqref{kern11} may also constitute  a starting point to construct a $q$-deformation for the determinantal point process associated with an $m$th Euclidean Landau level  or Ginibre-type point process in $\mathbb{C}$ as discussed by Shirai \cite{shirai}.\bigskip\\
\textbf{Remark. 5.1 } Note that the Stieltjes-Wigert polynomials (\cite{KS}, p.116) defined by 
\begin{equation}
s_n(x;q):=\sum_{k=0}^{n}\left[ 
\begin{array}{c}
n \\ 
k%
\end{array}%
\right] _{q}q^{k^2}x^{k}
\end{equation}
 are connected to  Rogers-Szeg\H {o} polynomials \eqref{RG-poly} by
\begin{equation}
s_n(x;q^{-1})=H_n(xq^{\frac{1}{2}-n};q).
\end{equation}
This  relation may also suggest the construction of another extension of $q$-deformed CS via the same
procedure, which would be attached to a suitable   Arik–Coon oscillator with $q > 1$ \cite{burb}.
\begin{center}
{\large\textbf{Appendix A : The wave function of the CS $\Psi_z^{q,m}$}}
\end{center}
By using \eqref{Eq:phizqm} and  replacing $\Phi_j^{m,q}(z)$ by their expressions in \eqref{Cjmq}, we need to establish a  closed form  for the following series
\begin{eqnarray*}\label{So31}
\mathcal{S}^m=\sum_{j\geq 0} \frac{(-1)^{m\wedge j} (q;q)_{m \vee j}q^{\binom{m\wedge j}{2}}\sqrt{1-q}^{|m-j|}|z|^{|m-j|}e^{-i(m-j)\theta}}{(q;q)_{|m-j|}\sqrt{q^{mj}(q;q)_m(q;q)_j}}\, P_{m\wedge j}\left( (1-q)z\bar{z};q^{|m-j|}|q\right)
  \varphi_j^{RS}(\xi).
\end{eqnarray*}
This sum can be cast into two quantities  as $S_{(<\infty)}^m+S_{(\infty)}^m$ where 
 \begin{eqnarray}
 \mathcal{S}_{(<\infty)}^m &=& \displaystyle\sum_{j=0}^{m-1} \frac{(-1)^{ j} (q;q)_{m }q^{\binom{ j}{2}}\sqrt{1-q}^{m-j}\bar{z}^{m-j}}{(q;q)_{m-j}\sqrt{q^{mj}(q;q)_m(q;q)_j}}P_{j}\left( (1-q)z\bar{z};q^{m-j}|q\right) \varphi_j^{RS}(\xi)\cr
 &-&  \displaystyle\sum_{j=0}^{m-1} \frac{(-1)^{m} (q;q)_{ j}q^{\binom{m}{2}}\sqrt{1-q}^{j-m}z^{j-m}}{(q;q)_{j-m}\sqrt{q^{mj}(q;q)_m(q;q)_j}}P_{m}\left( (1-q)z\bar{z};q^{j-m}|q\right) \varphi_j^{RS}(\xi),
 \end{eqnarray}
and
\begin{eqnarray}
S_{(\infty)}^m\left(z,\xi;q\right)=\frac{(-1)^mq^{({}^m_2)}(z\sqrt{1-q})^{-m}e^{-\frac{1}{2}\xi^2}}{\pi^{\frac{1}{4}}\sqrt{(q;q)_m}}\eta_{(\infty)}(m,z,\xi),
\end{eqnarray}
with  
\begin{eqnarray}
\label{etat}
\eta_{(\infty)}(m,z,\xi)=\sum_{j=0}^\infty\frac{(z\sqrt{1-q})^j}{\sqrt{q^{mj}}(q;q)_{j-m}} P_m((1-q)z\bar{z}; q^{j-m}|q)\;(i\sqrt{q})^jH_j\left(-e^{2i\kappa \xi};q\right).
\end{eqnarray}
Now, we apply  the relation (\cite{MoCa}, p.3) :  
\begin{equation}
 P_n(x;q^{-N}|q)=x^N(-1)^{-N}q^{\frac{N(N+1-2n)}{2}}\frac{(q^{N+1};q)_{n-N}}{(q^{1-N};q)_n} P_{n-N}(x;q^N|q)
 \end{equation}
for $N=j-m,n=j$ and $x=(1-q)z\bar{z}$, to obtain that $\mathcal{S}_{(<\infty)}^m=0.$ For the infinite sum, we first rewrite  the expression of the Wall polynomial  as (\cite{KS}, p.260) :
 \begin{equation}\label{wallreduce}
 P_n(x;a|q)=\frac{(x^{-1};q)_n}{(aq;q)_n} (-x)^n q^{-\binom{n}{2}}{}_2 \phi_1\left(\begin{matrix}q^{-n},0 \\ x q^{1-n} \end{matrix}\left|q;aq^{n+1}\right.\right),
 \end{equation}
in which we apply the identity 
\begin{equation}
(a;q)_{n}=(a^{-1}q^{1-n};q)_{n}(-a)^{n}q^{\binom{n}{2}}  \label{id15}
\end{equation}%
 to the term $(x^{-1};q)_n$. Therefore   \eqref{wallreduce} reduces to
\begin{equation}\label{wallreduce2}
P_n(x;a|q)=\frac{(xq^{1-n};q)_n}{(aq;q)_n}{}_2 \phi_1\left(\begin{matrix}q^{-n},0 \\ x q^{1-n} \end{matrix}\left|q;aq^{n+1}\right.\right).
\end{equation}
Setting  $n=m,\,x=(1-q)z\bar{z}$ and $a=q^{j-m}$ in \eqref{wallreduce2}, and appealing to the fact that
\begin{equation}
(a;q)_{n+k}=(a;q)_{n}(aq^{n};q)_{k},  \label{id14}
\end{equation}
the r.h.s of \eqref{etat} takes the form
\begin{eqnarray}
\eta_{(\infty)}(m,z,\xi)&=&(q^{1-m}\lambda;q)_m\sum_{j=0}^\infty\frac{y^j}{ (q;q)_{j}} H_j(\alpha;q) {}_2 \phi_1\left(\begin{matrix}q^{-m},0 \\\lambda q^{1-m} \end{matrix}\left|q;q^{1+j}\right.\right)
\end{eqnarray}
where $y=iz\sqrt{\frac{1-q}{q^{m-1}}}$ , $\alpha=-e^{2i\kappa \xi}$ and $\lambda=(1-q)z\bar{z}$. By recalling the definition of the  ${}_2 \phi_1$ series  and interchanging the order of summation, we get that 
\begin{eqnarray}\label{sumeta1}
\eta_{(\infty)}(m,z,\xi)
&=&(q^{1-m}\lambda;q)_m\sum_{k=0}^\infty\frac{(q^{-m};q)_k}{(\lambda q^{m-1};q)_k}\frac{q^k}{(q;q)_k}\sum_{j=0}^\infty\frac{(yq^k)^j}{ (q;q)_{j}} H_j(\alpha;q).
\end{eqnarray}
The last sum in \eqref{sumeta1} suggests us to make use  of the generating function of the Rogers-Szeg\H {o} polynomials (\cite{Ism29}, p.456):
\begin{equation}\label{RSGF}	
\displaystyle\sum_{j\geq 0}\frac{H_j(x;q) }{(q;q)_j}t^j=\frac{1}{(t;q)_{\infty}(q^{-1/2}xt;q)_{\infty}}
\end{equation} 
for $t=yq^k$ and $x=\alpha$. Therefore \eqref{sumeta1} also reads
\begin{eqnarray}
\label{sdetat}
\eta_{(\infty)}(m,z,\xi)= (q^{1-m}\lambda;q)_m\sum_{k=0}^\infty\frac{(q^{-m};q)_k}{(\lambda q^{m-1};q)_k}\frac{q^k}{(q;q)_k} \frac{1}{ (yq^k;q)_{\infty}\,(q^{-1/2}\alpha yq^k;q)_{\infty}}.
\end{eqnarray}
By applying the identity 
\begin{equation}
(a;q)_{\alpha }=\frac{(a;q)_{s}}{(aq^{s };q)_{\infty }}%
,\ aq^{s}\neq q^{-n},\ n\in \mathbb{N}, \label{id11}
\end{equation}
Eq.\eqref{sdetat}  transforms to 
\begin{eqnarray}
\label{qetat}
 \eta_{(\infty)}(m,z,\xi)=\frac{(q^{1-m}\lambda;q)_m}{(y ,q^{-1/2}\alpha y ;q)_\infty}\sum_{k=0}^\infty\frac{(q^{-m},q^{-1/2}\alpha y, y;q)_k}{(\lambda q^{m-1};q)_k}\frac{q^k}{(q;q)_k}
\end{eqnarray}
where the last sum may be expressed in terms of a  ${}_3 \phi_2$ series  as
\begin{eqnarray}
\label{qqqetat}
 \eta_{(\infty)}(m,z,\xi)=\frac{(q^{1-m}\lambda;q)_m}{(y ,q^{-1/2}\alpha y ;q)_\infty}{}_3 \phi_2\left(\begin{matrix}q^{-m},y,q^{-1/2}\alpha y \\ \lambda q^{1-m},0 \end{matrix}\Big|q;q\right)
\end{eqnarray}
with   $$y=iz\sqrt{\frac{1-q}{q^{m-1}}}=q^{-1/4}e^{i\kappa \xi}z\sqrt{\frac{1-q}{q^{m-1}}}\; iq^{1/4}e^{-i\kappa \xi} $$
and 
  $$q^{-1/2}\alpha y=-iz\sqrt{\frac{1-q}{q^{m-1}}}e^{2i\kappa \xi}=q^{-1/4}e^{i\kappa \xi}z\sqrt{\frac{1-q}{q^{m-1}}}\; \frac{1}{iq^{1/4}e^{-i\kappa \xi}}.$$
  Finally, recalling the definition of  Al-Salam-Chihara polynomials  \eqref{SaChidef} and taking into account the previous prefactors, we arrive at the announced result  \eqref{SalCHi}.  $\square$ \medskip\\
  \begin{center}
{\large\textbf{Appendix B : $\mathcal{B}_{m}^{(q)}$ at the limit $q\to 1$}}
\end{center}
We first use the identity \eqref{qfactor} to rewrite \eqref{CSBMQ} as  
\begin{equation*}
\mathcal{B}_{m}^{(q)}[f](z)=(-1)^m\left(q^{\frac{2m^2-m}{2}}2^m[m]_q!\sqrt{\pi}\right)^{-1/2} 
\end{equation*}
\begin{equation}\label{Bmqlimit}
\times \int_{\mathbb{R}}\frac{e^{im\kappa \xi-\frac{1}{2}\xi^2} }{(iz\sqrt{\tfrac{1-q}{q^{m-1}}},-iz \sqrt{\tfrac{1-q}{q^m}} e^{2i\kappa \xi};q)_\infty}\,\frac{ Q_m\left(\tau s;2\tau \alpha,2\tau \beta|q  \right)}{\tau^m}
 f(\xi)d\xi
\end{equation}
where 
\begin{equation}\label{paramsalam}
s=\sqrt{\frac{q^{m-1}}{1-q}}\frac{ie^{-i\kappa\xi}q^{1/4}-ie^{i\kappa\xi}q^{-1/4}}{\sqrt{2}},\;\alpha=q^{-1/4}\frac{z}{\sqrt{2}}e^{\frac{i\kappa \xi}{2}},\;\beta=q^{1/4}\frac{\bar{z}}{\sqrt{2}}e^{\frac{-i\kappa \xi}{2}}\: \mathrm{and}\: \tau=\sqrt{\frac{1-q}{2q^{m-1}}}.
\end{equation}
Denoting 
\begin{equation}
G_q(z;\xi):= \frac{1}{\left(iz\sqrt{\tfrac{1-q}{q^{m-1}}},-iz \sqrt{\tfrac{1-q}{q^m}} e^{2i\kappa \xi};q\right)_{\infty}}
\end{equation}
and using  \eqref{id11}, it  follows  that
\begin{eqnarray}
\mathrm{Log}\,G_q(z;\xi)&=& -\sum_{k\geq 0} \mathrm{Log}\left(1+izq^{\frac{-m}{2}}\sqrt{1-q}(e^{2i\kappa \xi}-\sqrt{q}) q^k+z^2(1-q)q^{\frac{1-m}{2}}e^{2i\kappa x}q^{2k}\right)\cr
&=& -izq^{\frac{-m}{2}}\sqrt{1-q}(e^{2i\kappa \xi}-\sqrt{q}) \sum_{k\geq 0} q^k-z^2(1-q)q^{\frac{1-m}{2}}e^{2i\kappa \xi} \sum_{k\geq 0} q^{2k}+{o}(1-q)\cr
&=& izq^{\frac{-m}{2}}(\sqrt{q}-e^{2i\kappa \xi})\frac{1}{\sqrt{1-q}}-z^2q^{\frac{1-m}{2}}e^{2i\kappa \xi}\frac{1}{1+q}+{o}(1-q)
\end{eqnarray}
where the quantities $\frac{{i}zq^{\frac{-m}{2}}(\sqrt{q}-e^{2i\kappa \xi})}{\sqrt{1-q}}$ and $z^2q^{\frac{1-m}{2}}e^{2i\kappa \xi}\frac{1}{1+q}$
go  to $\sqrt{2}\xi z$ and $\frac{1}{2}z^2$ as $q\to 1$ (or $\kappa \to 0)$ respectively. Therefore $G_q(z;\xi)\to exp(\sqrt{2}\xi z-\frac{1}{2}z^{2})$ as $q\to 1$.
By another side, we may apply the result  (\cite{AtAt}, p.3):
\begin{equation}
\lim_{q\rightarrow 1} \tau^{-m}Q_m(\tau s;2\tau \alpha,2\tau \beta|q)=H_m(s-\alpha-\beta)
\end{equation}
for the parameters $s,\;\alpha,\;\beta$ and $\tau$ as in \eqref{paramsalam}, to arrive at 
$$\lim_{q\rightarrow 1}(-1)^m\left(q^{\frac{2m^2-m}{2}}2^m[m]_q!\sqrt{\pi}\right)^{-1/2}Q_m\left(\tau s;2\tau \alpha,2\tau \beta|q  \right)=(-1)^m(2^{m}m!\sqrt{\pi})^{-\tfrac{1}{2}}H_{m}\left( \xi -\frac{z+\bar{z}}{\sqrt{2}}\right).$$
Finally, summarizing the above calculations we complete the proof of   the assertion  \eqref{tr1}. $\square$\medskip\\
\begin{center}
\textbf{Appendix C : The reproducing kernel $K_{q,m}(z,w)$ and its limit at  $q\to 1$}
\end{center}
Since the  $\Phi_j^{m,q}$'s form  an orthonormal basis of the space $\mathcal{A}_{q,m}^2(\mathbb{C})$, the corresponding reproducing kernel is given, according the general theory \cite{Aro50, Saitoh} by
\begin{equation}\label{RKdem1}
K_{q,m}(z,w):=\sum_{j=0}^\infty \Phi_j^{q,m}(z) \overline{\Phi_j^{q,m}(w)}.
\end{equation}
For that we replace the   $\Phi_j^{q,m}$'s by their  expressions   in    \eqref{Cjmq} and we split the sum \eqref{RKdem1} into two parts as 
 \begin{equation}\label{prop1eq01}
K_{q,m}(z,w)=\mathcal{S}_{(<\infty)}^m(z,w;q)+S_{(\infty)}^m\left(z,w;q\right)
 \end{equation}
 where
\begin{eqnarray*}
 \mathcal{S}_{(<\infty)}^m(z,w;q)&=& \displaystyle\sum_{j=0}^{m-1} \frac{(q,q;q)_{m }q^{2\binom{ j}{2}}{(1-q)}^{m-j}\bar{z}^{m-j}w^{m-j}}{(q,q;q)_{m-j}{q^{mj}(q;q)_m(q;q)_j}}P_{j}\left( (1-q)z\bar{z};q^{m-j}|q\right)P_{j}\left( (1-q)w\bar{w};q^{m-j}|q\right)\cr
 &-&  \displaystyle\sum_{j=0}^{m-1} \frac{(q,q;q)_{j}q^{2\binom{ m}{2}}{(1-q)}^{j-m}{z}^{j-m}\bar{w}^{j-m}}{(q,q;q)_{j-m}{q^{mj}(q;q)_m(q;q)_j}}P_{m}\left( (1-q)z\bar{z};q^{j-m}|q\right)P_{m}\left((1-q)w\bar{w};q^{j-m}|q\right),
 \end{eqnarray*}
and
\begin{eqnarray}\label{finisum}
S_{(\infty)}^m\left(z,w;q\right)&=&\displaystyle\sum_{j \geq 0} \frac{(q,q;q)_{j}q^{2\binom{ m}{2}}{(1-q)}^{j-m}{z}^{j-m}\bar{w}^{j-m}}{(q,q;q)_{j-m}{q^{mj}(q;q)_m(q;q)_j}}P_{m}\left( (1-q)z\bar{z};q^{j-m}|q\right)P_{m}\left((1-q)w\bar{w};q^{j-m}|q\right))\cr
&=&\frac{q^{2\binom{m}{2}}}{(q;q)_m}\left(\frac{1}{\lambda}\right)^m\sum_{j\geq 0}\frac{(q;q)_j}{(q,q;q)_{j-m}}\left(\frac{\lambda}{q^m}\right)^jP_m\left( \alpha;q^{j-m}|q\right)P_m\left( \beta;q^{j-m}|q\right),
\end{eqnarray}
where $\lambda=(1-q)z\bar{w},\,\alpha=(1-q)z\bar{z}$ and $\beta=(1-q)w\bar{w}$. By making use of  the relation (\cite{MoCa}, p.3):
\begin{equation}
 P_n(x;q^{-N}|q)=x^N(-1)^{-N}q^{\frac{N(N+1-2n)}{2}}\frac{(q^{N+1};q)_{n-N}}{(q^{1-N};q)_n} P_{n-N}(x;q^N|q)
 \end{equation}
 for  parameters $N=j-m,n=j$, $x=\alpha$ and $x=\beta$, we obtain that  $\mathcal{S}_{(<\infty)}^m(z,w;q)=0.$ For the infinite sum  \eqref{finisum}, we rewrite  the Wall polynomial  as (\cite{KS}, p.260):
 \begin{equation}
 P_n(x;a|q)=\frac{(x^{-1};q)_n}{(aq;q)_n} (-x)^n q^{-\binom{n}{2}}{}_2 \phi_1\left(\begin{matrix}q^{-n},0 \\ x q^{1-n} \end{matrix}\left|q;aq^{n+1}\right.\right)
 \end{equation}
with $n=m,\,x=(1-q)\lambda$ and $a=q^{j-m}$. Next, by using  \eqref{id15}-\eqref{id14} respectively, Eq.\eqref{prop1eq01} takes the form
\begin{eqnarray}
K_{q,m}(z,w) &=& \frac{q^{2\binom{m}{2}}}{(q;q)_m}\left(\frac{1}{\lambda}\right)^m (q^{1-m}\alpha;q)_m(q^{1-m}\beta;q)_m \mathsf{S}_q^m(\alpha;\beta) 
\end{eqnarray} 
where 
\begin{equation}\label{1.7}
\mathsf{S}_q^m(\alpha;\beta) :=\sum_{j \geq 0} \frac{ (q;q)_j}{(q;q)_{j-m}^2 }\frac{1}{(q^{j-m+1};q)_m^2}\left(\frac{\lambda}{q^m}\right)^j\,{}_2\phi_1\left(\begin{array}{c}q^{-m}, 0\\
\alpha q^{1-m}\end{array}\Big|q; q^{j+1}\right){}_2\phi_1\left(\begin{array}{c}q^{-m}, 0\\
\beta q^{1-m}\end{array}\Big|q; q^{j+1}\right).
\end{equation}
Using the identity  $(q;q)_{j-m}^2(q^{j-m+1};q)_m^2=(q;q)_j^2$, we  may also write 
\begin{eqnarray}\label{eq12}
\mathsf{S}_q^m(\alpha;\beta) &=&  \sum_{j \geq 0} \frac{(q^{-m}\lambda)^j}{(q;q)_j}\,\sum_{k \geq 0} \frac{(q^{-m};q)_k}{(\alpha q^{1-m};q)_k}\,\frac{(q^{j+1})^k}{(q;q)_k} \sum_{l \geq 0} \frac{(q^{-m};q)_l}{(\beta q^{1-m};q)_l}\,\frac{(q^{j+1})^l}{(q;q)_l}\cr
&=& \sum_{k,l \geq 0} \frac{(q^{-m};q)_k}{(\alpha q^{1-m};q)_k}\,\frac{q^k}{(q;q)_k}  \frac{(q^{-m};q)_l}{(\beta q^{1-m};q)_l}\,\frac{q^l}{(q;q)_l}\,\sum_{j \geq 0} \frac{(q^{-m+k+l}\lambda)^j}{(q;q)_j}.
\end{eqnarray}
Now, by applying the $q$-binomial theorem 
\begin{equation}\label{binothe}
\sum_{n \geq 0} \frac{(a;q)_n}{(q;q)_n}\xi^n=\frac{(a\xi;q)_{\infty}}{(\xi;q)_{\infty}},\quad |\xi|<1
\end{equation}
 for $\xi=q^{-m+k+l}\lambda$ and $a=0$, the last sum reads
\begin{eqnarray}\label{eq13}
\mathsf{S}_q^m(\alpha;\beta)&=& \sum_{k,l \geq 0} \frac{(q^{-m};q)_k}{(\alpha q^{1-m};q)_k}\,\frac{q^k}{(q;q)_k}  \frac{(q^{-m};q)_l}{(\beta q^{1-m};q)_l}\,\frac{q^l}{(q;q)_l}\,\frac{1}{(q^{-m+k+l}\lambda;q)_{\infty}}.
\end{eqnarray}
By applying the identity \eqref{id11} to the factor $\frac{1}{(q^{-m+k+l}\lambda;q)_\infty}$, Eq.\eqref{eq13} can be rewritten as
\begin{equation}\label{eq14}
\mathsf{S}_q^m(\alpha;\beta)=\frac{1}{(q^{-m}\lambda;q)_{\infty}}\sum_{k\geq 0} \frac{(q^{-m};q)_k}{(\alpha q^{1-m};q)_k}\,\frac{q^k}{(q;q)_k}\sum_{l\geq 0}  \frac{(q^{-m};q)_l(q^{-m}\lambda;q)_{k+l}}{(\beta q^{1-m};q)_l}\,\frac{q^l}{(q;q)_l}.
\end{equation}
Next, by writing $(q^{-m}\lambda;q)_{l+k}=(q^{-m}\lambda;q)_k(q^{k-m}\lambda;q)_l$, Eq.\eqref{eq14} transforms to  
\begin{eqnarray}\label{eq132}
\mathsf{S}_q^m(\alpha;\beta)&=& \frac{1}{(q^{-m}\lambda;q)_{\infty}}\sum_{k\geq 0} \frac{(q^{-m},q^{-m}\lambda;q)_k}{(\alpha q^{1-m};q)_k}\,\frac{q^k}{(q;q)_k}\sum_{l\geq 0}  \frac{(q^{-m},q^{k-m}\lambda;q)_l}{(\beta q^{1-m};q)_l}\,\frac{q^l}{(q;q)_l}\cr
&=& \frac{1}{(q^{-m}\lambda;q)_{\infty}}\sum_{k\geq 0} \frac{(q^{-m},q^{-m}\lambda;q)_k}{(\alpha q^{1-m};q)_k}\,\frac{q^k}{(q;q)_k} {}_2\phi_1\left(\begin{array}{c}q^{-m}, q^{k-m}\lambda\\
\beta q^{1-m}\end{array}\Big|q; q\right),
\end{eqnarray}
in terms of the  ${}_2\phi_1$ series. The latter one satisfies the identity (\cite{GR}, p.10):
\begin{equation}\label{21 ident}
{}_2\phi_1\left(\begin{array}{c}q^{-n}, b\\
c\end{array}\Big|q; q\right) =\frac{(b^{-1}c;q)_n}{(c;q)_n}b^n,\quad n=0,1,2,....
\end{equation} 
Setting  $n=m$, $b=q^{k-m}\lambda$ and $c=\beta q^{1-m}$ in \eqref{21 ident} allows us to rewrite \eqref{eq132} as
\begin{eqnarray}
\mathsf{S}_q^m(\alpha;\beta) &=& \frac{1}{(q^{-m}\lambda;q)_{\infty}}\sum_{k\geq 0} \frac{(q^{-m},q^{-m}\lambda;q)_k}{(\alpha q^{1-m};q)_k}\,\frac{q^k}{(q;q)_k}\,\frac{(q^{1-k}\frac{\beta}{\lambda};q)_m}{(q^{1-m}\beta;q)_m}(q^{k-m}\lambda)^m.
\end{eqnarray} 
Furthermore, applying the identity 
\begin{equation}
(aq^{-n};q)_{k}=\frac{(a^{-1}q;q)_{n}}{(a^{-1}q^{1-k};q)_{n}}%
(a;q)_{k}q^{-nk},\quad a\neq 0  \label{id16}
\end{equation}
to $(q^{1-k}\frac{\beta}{\lambda};q)_m$, leads to  
\begin{eqnarray}
\mathsf{S}_q^m(\alpha;\beta) &=& \frac{(q^{-m}\lambda)^m(q\frac{{w}}{{z}};q)_m}{(q^{1-m}\beta;q)_m(q^{-m}\lambda;q)_{\infty}}\sum_{k\geq 0} \frac{(q^{-m},q^{-m}\lambda,\frac{{z}}{{w}};q)_k}{(q^{1-m}\alpha,q^{-m}\frac{{z}}{{w}} ;q)_k}\,\frac{q^k}{(q;q)_k}.
\end{eqnarray} 
We now summarize the above calculations as follows 
\begin{eqnarray}\label{RKL2}
K_{q,m}(z,w) &=& \frac{q^{-m}(q^{1-m}\alpha;q)_m (q\frac{{w}}{{z}};q)_m}{(q^{-m}\lambda;q)_{\infty}(q;q)_m}\sum_{k\geq 0} \frac{(q^{-m},q^{-m}\lambda,\frac{{z}}{{w}};q)_k}{(q^{1-m}\alpha,q^{-m}\frac{{z}}{{w}} ;q)_k}\,\frac{q^k}{(q;q)_k}\cr
&=& \frac{(q^{1-m}\alpha;q)_m (q\frac{{w}}{{z}};q)_m}{q^m(q^{-m}\lambda;q)_{\infty}(q;q)_m}\,{}_3\phi_2\left(\begin{array}{c}q^{-m}, q^{-m}\lambda,\frac{{z}}{{w}}\\
q^{1-m}\alpha,q^{-m}\frac{z}{w}\end{array}\Big|q; q\right).
\end{eqnarray}
By making appeal  to the finite Heine transformation (\cite{GA09}, p.2): 
 \begin{equation}
 {}_3\phi_2\left(\begin{array}{c}q^{-n},\xi,\beta\\
\gamma,q^{1-n}/\tau\end{array}\Big|q; q\right)=\frac{(\xi\,\tau;q)_n}{(\tau;q)_n}\;{}_3\phi_2\left(\begin{array}{c}q^{-n},\gamma/\beta,\xi\\
\gamma,\xi\,\tau\end{array}\Big|q; \beta\,\tau q^n\right)
 \end{equation}
 for the parameters $\xi=z/w,\;\beta=q^{-m}\lambda,\;\gamma=q^{1-m}\alpha,\; \tau=qw/z$, we obtain
\begin{eqnarray}
\hspace*{2cm} K_{q,m}(z,w)=\frac{(q^{1-m}(1-q)z\bar{z};q)_{m}}{ q^m( q^{-m}(1-q)z\bar{w};q)_\infty}
{}_3\phi_2\left(\begin{array}{c}q^{-m},\frac{{z}}{{w}},q\frac{\bar{z}}{\bar{w}}\\
 q,q^{-m+1}(1-q)z\bar{z}\end{array}\Big|q;q(1-q)w\bar{w}\right).
\end{eqnarray}
For the limit of $K_{q,m}(z,w)$ as $q\to 1$, we may use  \eqref{qexpo1} to see that  
\begin{equation}
\lim_{q\to1}\frac{(q^{1-m}(1-q)z\bar{z};q)_{m}}{ q^m( q^{-m}(1-q)z\bar{w};q)_\infty}=\lim_{q\to1}q^{-m}(q^{1-m}(1-q)z\bar{z};q)_{m}\;e_q(q^{-m}z\bar{w})=\mathrm{exp}\left(z\bar{w}\right).
\end{equation}
By another side, we observe that  $(q^{-m};q)_k=0, \forall k>m$, meaning that  ${}_3\phi_2$ series  in  \eqref{kern11} terminates as 
\begin{eqnarray}\label{terseri}
\sum_{k=0}^{m}\frac{(q^{-m},z/w,q\bar{z}/\bar{w};q)_k}{(q^{1-m}(1-q)z\bar{z},q;q)_k}\,\frac{( q(1-q)w\bar{w})^k}{(q;q)_k}.
\end{eqnarray}
Thus, from the identity 
\begin{equation}
\begin{bmatrix}
s \\ 
k%
\end{bmatrix}%
_{q}:=(-1)^{k}q^{k s -\binom{k}{2}}\frac{(q^{-{s }};q)_{k}}{%
(q;q)_{k}},\:s \in \mathbb{C}  \label{binomger}
\end{equation}
together with \eqref{qfactor} we, successively, have 
\begin{eqnarray}\label{limgfunc}
\lim_{q\to 1}\sum_{k=0}^{m}\frac{(q^{-m},z/w,q\bar{z}/\bar{w};q)_k}{(q^{1-m}(1-q)z\bar{z},q;q)_k}\,\frac{( q(1-q)w\bar{w})^k}{(q;q)_k}&=&\sum_{k=0}^{m}\lim_{q\to 1}\left(\frac{(q^{-m};q)_k}{(q;q)_k}\frac{(z/w,q\bar{z}/\bar{w};q)_k}{(q^{1-m}(1-q)z\bar{z};q)_k}\,\frac{(1-q)^k}{(q;q)_k}\,q^{k}(w\bar{w})^k\right)\cr
&=&\sum_{k=0}^{m}\lim_{q\to 1}\left(\begin{bmatrix} m\\ k \end{bmatrix}_q (-1)^kq^{\binom{k}{2}-mk}\frac{(z/w,q\bar{z}/\bar{w};q)_k}{(q^{1-m}(1-q)z\bar{z};q)_k}\,\frac{q^{k}(w\bar{w)}^k}{[k]_q!}\right)\cr
&=&\displaystyle\sum_{k=0}^m (-1)^k \binom{m}{k} \frac{|z-w|^{2k}}{k!}.
\end{eqnarray}
Finally, by noticing that  the last  sum in \eqref{limgfunc} is the evaluation of the Laguerre polynomial $L_m^{(0)}$ at  $|z-w|^{2}$ (\cite{KS}, p.108), the proof is completed. $\square$\bigskip\\
\textbf{Acknowledgements}

The authors would like to thank the anonymous referees for their valuable
comments and suggestions that improved the quality of this paper. The authors also are
thankful to the Moroccan Association of Harmonic Analysis and Spectral Geometry.

\end{document}